\documentclass[twocolumn]{aastex62}
\usepackage[utf8]{inputenc}
\usepackage{longtable}
\usepackage{graphicx}
\usepackage{amsmath,amssymb}
\usepackage{color}
\usepackage{units}
\usepackage{epstopdf}
\usepackage{hyperref}
\usepackage{multirow}
\usepackage{url}
\usepackage{xcolor}
\usepackage[caption=false]{subfig}
\usepackage{rotating}

\newcommand{\beq}{\begin{equation}}
\newcommand{\eeq}{\end{equation}}
\newcommand{\bdm}{\begin{displaymath}}
\newcommand{\edm}{\end{displaymath}}

\definecolor{Gray}{gray}{0.9}
\definecolor{orange}{rgb}{0.9,0.5,0}

\graphicspath{{./plots/}}

\begin{document}

\title{Dynamic Scheduling: Target of Opportunity Observations of Gravitational Wave Events}

\author{Mouza Almualla}
\affil{American University of Sharjah, Physics Department, PO Box 26666, Sharjah, UAE}

\author[0000-0002-8262-2924]{Michael W. Coughlin}
\affil{School of Physics and Astronomy, University of Minnesota, Minneapolis, Minnesota 55455, USA}
\affil{Division of Physics, Mathematics, and Astronomy, California Institute of Technology, Pasadena, CA 91125, USA}

\author{Shreya Anand}
\affil{Division of Physics, Mathematics, and Astronomy, California Institute of Technology, Pasadena, CA 91125, USA}

\author{Khalid Alqassimi}
\affil{American University of Sharjah, Physics Department, PO Box 26666, Sharjah, UAE}

\author{Nidhal Guessoum}
\affil{American University of Sharjah, Physics Department, PO Box 26666, Sharjah, UAE}

\author[0000-0001-9898-5597]{Leo P. Singer}
\affiliation{Astrophysics Science Division, NASA Goddard Space Flight Center, MC 661, Greenbelt, MD 20771, USA}
\affiliation{Joint Space-Science Institute, University of Maryland, College Park, MD 20742, USA}

\begin{abstract}
The simultaneous detection of electromagnetic and
gravitational waves from the coalescence of two neutron stars
(GW170817 and GRB170817A) has ushered in a new era of ``multi-messenger''
astronomy, with electromagnetic detections spanning from gamma to radio. This great
opportunity for new scientific investigations raises the issue of how
the available multi-messenger tools can best be integrated to
constitute a powerful method to study the transient universe in
particular. To facilitate the classification of possible optical counterparts to
gravitational-wave events, it is important to optimize the scheduling
of observations and the filtering of transients, both key elements of
the follow-up process. In this work, we describe the existing workflow
whereby telescope networks such as GRANDMA and GROWTH are currently scheduled;
we then present modifications we have developed for the scheduling
process specifically, so as to face the relevant challenges that have
appeared during the latest observing run of Advanced LIGO and Advanced Virgo. We address issues with scheduling more than one epoch for multiple fields within a skymap, especially for large and disjointed localizations. 
This is done in two ways: by optimizing the maximum number of fields that can be scheduled, and by splitting up the lobes within the skymap by right ascension to be scheduled individually. 
In addition, we implement the ability to take previously observed fields into consideration when rescheduling. 
We show the improvements that these modifications produce in making the search for optical counterparts more efficient, and we point to areas needing further improvement.

\end{abstract}


\section{Introduction}
\begin{figure*}[t]
    \includegraphics[width=3.5in]{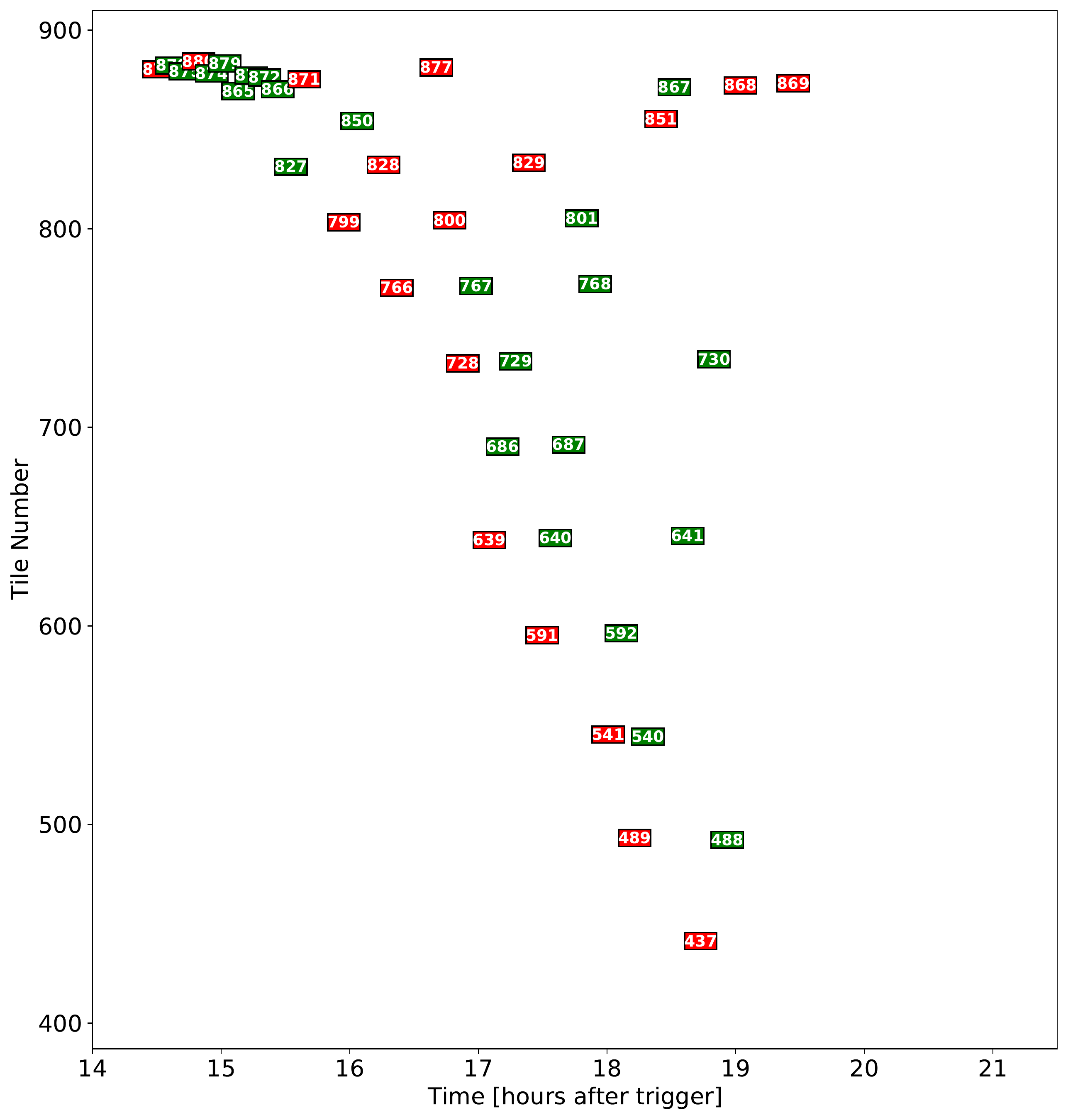}
    \includegraphics[width=3.5in]{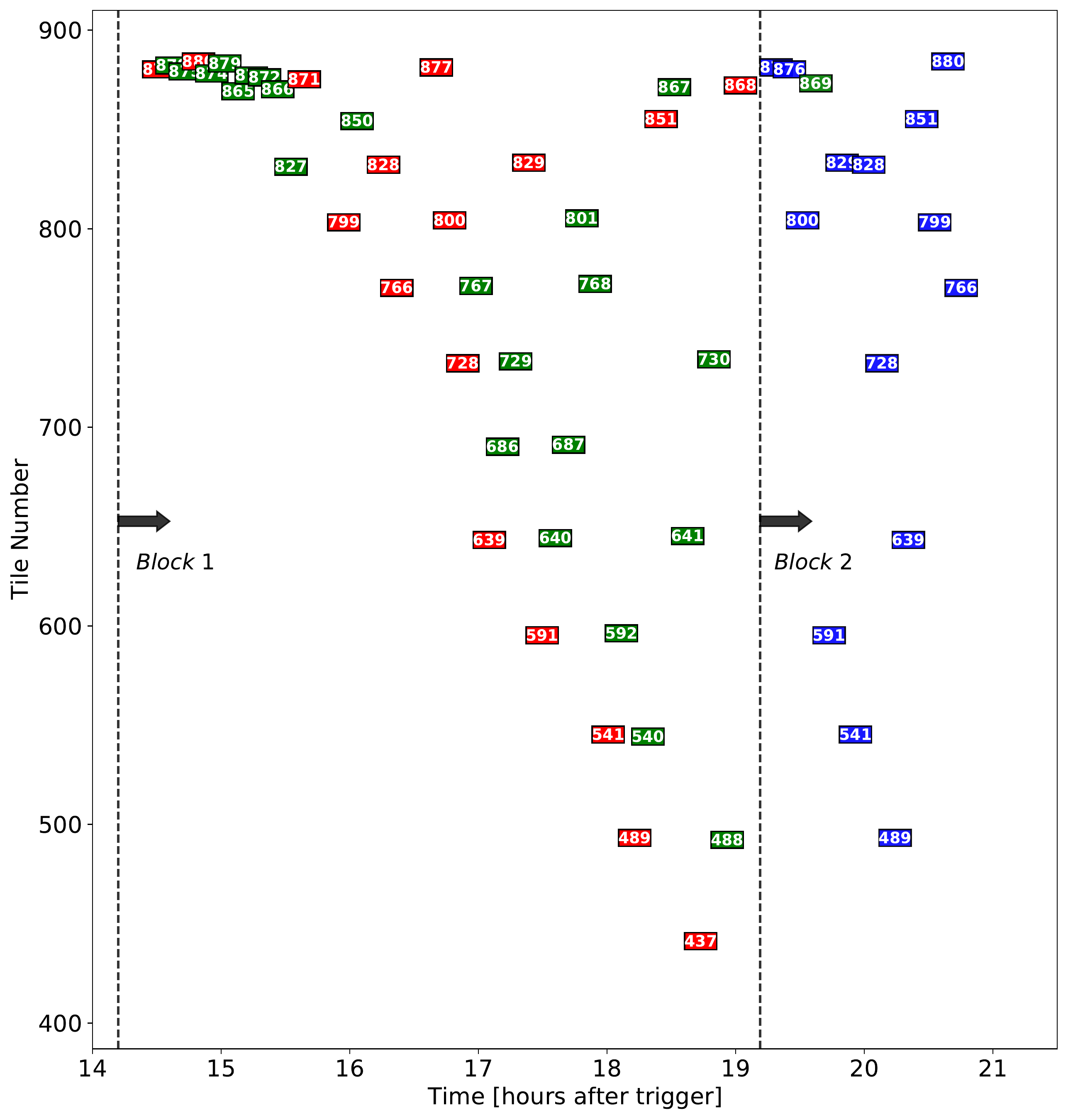}
    \label{fig:normvssupersched}
    \caption{Plots of coverage for S190426c \citep{ChEA2019a} without (left) and with (right) the use of the SuperScheduler algorithm. Red indicates that the corresponding field could not be observed during that respective round, and green indicates that the observation was successful. For simplicity, all fields were scheduled in the same filter for both of the cases shown. Breaking the night up into two blocks and using the SuperScheduler, 14 previously failed attempts at observation were successfully rescheduled (shown in blue). The rectangle widths (representing exposure time) have been scaled by a factor of 3 for visualization purposes, and the respective tile number is labelled at the center of each rectangle.}
\end{figure*}

The first and second observing runs of the global network of gravitational wave (GW) interferometers, comprising the Advanced Virgo \citep{adVirgo} and the twin Advanced LIGO \citep{aLIGO} detectors, yielded the detection of a total of ten binary black hole (BBH) mergers and one binary neutron star (BNS) coalescence \citep{AbEA2018b}. Most recently, the improved sensitivity of the instruments during the third observing run (O3) has resulted in 56 GW candidates - many of which have been classified as BNS or neutron star-black hole (NSBH) mergers (information about these candidates can be found on the Gravitational-wave Candidate Event Database, or GraceDB\footnote{\url{https://gracedb.ligo.org/}}).

Due to the association of BNS and NSBH mergers with potentially detectable electromagnetic counterparts \citep{MeBe2012, nakar2019electromagnetic}, substantial efforts have been invested into optimizing follow-up observations of such candidates \citep[e.g.,][]{CoAh2019b, GoAn2019, GoHo2019, Andreoni2020}. These counterparts may come in the form of short gamma-ray bursts (sGRBs) accompanied by optical and NIR transients (``kilonovae'', or KNe) powered by the decay of r-process nuclei that are synthesized in the merger ejecta, as well as prolonged radio emission resulting from the interaction of the sub-relativistic ejecta with the surrounding medium \citep[e.g.,][]{LiPa1998,NaPi2011,MeBe2012,PiNa2013,Tanaka2016,GoVe2017,Savchenko_2017,GuZi2018,Metzger2019}.

The culmination of these follow-up efforts came to fruition on the 17th of August, 2017, unveiling the new era of multi-messenger astronomy with the detection of GW170817 along with the short gamma-ray burst GRB 170817A \citep{AbEA2017e}, and multiple independent discoveries of the optical transient counterpart AT 2017gfo in NGC 4993 (D $\sim40$ Mpc) by various teams (see \cite{AbEA2017h} and references therein), the first of which was announced by \cite{2017Sci...358.1556C}. The three Advanced LIGO and Virgo instruments had detected a signal that was determined to have likely originated from a BNS coalescence; the source was well-constrained, initially localized to $\sim 31$ deg$^2$ at the 90\% credibility level and with luminosity distance $40\pm8$ Mpc \citep{gcn21513,AbEA2017b}. The unprecedented nature of these detections has since led to such scientific gains as the ability to probe into the workings of r-process nucleosynthesis in kilonovae \citep[e.g.,][]{ChBe2017,2017Sci...358.1556C, CoBe2017,PiDa2017,SmCh2017,WaHa2019,KaKa2019} and the expansion rate of the Universe \citep{2017Natur.551...85A,HoNa2018,CoDi2019}, as well as constrain properties of neutron stars such as mass, radius, and tidal deformability in novel ways \citep[e.g.,][]{BaJu2017, MaMe2017, CoDi2018b, CoDi2018, CoDi2019b, AnEe2018, MoWe2018,RaPe2018,AbAb2018,Lai2019}.

The majority of candidate BNS and NSBH mergers during O3 had localization areas leaning towards the thousands of square degrees, with the exception of probable NSBH merger S190814bv (which has an updated 90\% credible region of 23 square degrees \citep{gcn25333}; these numbers are in stark contrast with the aforementioned localization area of GW170817. The sky localization is expected to improve in subsequent observing runs, thanks to the addition of other detectors to the GW network \citep{AbEA2020}; however, there will still likely be events with localization areas of the order of hundreds of square degrees, which is much larger than the field-of-view of most electromagnetic facilities, and so there will continue to be challenges in obtaining significant coverage of the skymap in the future. It is thus important to optimize our methods in performing follow-ups to GW triggers, which will greatly increase the odds of detecting an electromagnetic counterpart.

A codebase named \texttt{gwemopt}\footnote{\url{https://github.com/mcoughlin/gwemopt}} (Gravitational-Wave ElectroMagnetic OPTimization) was hence developed \citep{CoTo2018}, aimed at optimizing the scheduling of Target of Opportunity (ToO) telescope observations immediately after a GW detection. This code breaks the process down into three parts: tiling, time allocation, and scheduling. During the tiling step, it takes the HEALPix GW skymap and splits it up into ``tiles'' according to the FOV characteristics of the given telescope. It then goes on to allocate time to the tiles that are available for observation, which is dependent on the algorithm that is utilized for the plan. \texttt{gwemopt} finally proceeds to schedule these observations, taking into account factors such as the probability associated with the tiles, slew time, and observability. One way to further optimize the follow-up process is through the implementation of network-level telescope observations during scheduling \citep[this is discussed in-depth, for example, in][]{CoAn2019}, in which various telescopes around the world work together to achieve maximum coverage of the localization area for a given event. This is an especially relevant issue in the case of ToO observations, as multi-telescope observations will improve our ability to cover areas in the localization that may not be accessible to one given telescope (e.g., the localization area could extend into different hemispheres); in addition, this will allow different telescopes to coordinate in imaging the same patch of the sky in different filters and perform independent visits separated in time. 

In this paper, we delineate the new additions to \texttt{gwemopt} that build upon these ideas and expand on the currently available features. These features will facilitate the scheduling process in the case of both multi- and single- telescope observations. In Section~\ref{sec:dynamicsched}, we discuss the novel ability for \texttt{gwemopt} to take into account previously completed observations when re-scheduling, and in Section~\ref{sec:filterbalancing}, we describe two features that drastically improve multi-epoch coverage of events. In Section~\ref{sec:conclusion}, we conclude by discussing the role that these features play in the broader context of large telescope networks.

\section{The SuperScheduler Algorithm}
\label{sec:dynamicsched}

Although various factors such as observability and telescope location are taken into account during the scheduling process, light pollution, bad weather conditions, and unanticipated telescope-related failures may often lead to unsuccessful attempts at observation. When scheduling or re-scheduling these observations, \texttt{gwemopt} does not have any information as to whether a given tile has already been observed or not. This limitation poses some problems, since there is a possibility that the \texttt{gwemopt} pipeline will schedule tiles that were already observed rather than prioritizing unobserved tiles and increasing coverage of the localization.

This is an especially important point to consider in the case of multi-telescope observations, as there should be a way to schedule different telescopes and take previous observation rounds into account. The SuperScheduler can do this by going through a given number of iterations of the scheduling process, with each iteration corresponding to an observation round. The algorithm is able to take previous rounds into account when rescheduling by reading in information about which tiles have or have not been observed, it then sets the 2D spatial probability of the GW skymap enclosed in the observed tiles to 0 before the next round is scheduled.

This step improves the efficiency of the scheduling process since \texttt{gwemopt} no longer redundantly schedules the same tiles for re-observation. The algorithm can work for multiple telescopes in each round, and the telescopes can also be changed between different iterations. In cases where observations in more than one filter are scheduled, the SuperScheduler also takes the filter in which the field was observed into account. So if a given field has only been observed in the \textit{g}-band, for example, it can still schedule a second exposure in the \textit{r}-band the next time around rather than completely ignoring the field.

In order to test the capabilities of the SuperScheduler algorithm, we performed a simulation using the BNS merger candidate S190426c \citep{ChEA2019a} in which a 50\% failure rate (exaggerated for purposes of demonstration) was assumed for the attempted observations. The performance between two different cases was compared; in the first case, the whole night was scheduled normally, and in the second case, the night was broken down into two blocks, and failed observations in the first block were taken into account when scheduling the second block. As shown on the left of Figure~\ref{fig:normvssupersched}, the normal scheduling algorithm does not take into account whether certain tiles have had successful (shown in green) or unsuccessful (shown in red) observation attempts throughout the scheduling process. As a result, there are no previously unobserved tiles scheduled for observation (which would be shown in blue). Conversely, the results using the SuperScheduler algorithm on the right of Figure~\ref{fig:normvssupersched} show that, breaking the night down into two blocks, prioritizing tiles that were not successfully observed in the first block (as indicated in the figure) led to almost all of the failed observations being re-scheduled in the second block. We also note that many more fields are scheduled past the starting point of the second block (around 19 hours after trigger) when using the SuperScheduler; this is because the algorithm allows the scheduler to revisit the fields that failed to be observed in the first round. This ability is not available when using the normal scheduling algorithm, and most of the localization had already been covered by that point in time, thus leading to very few additional fields scheduled past that point in comparison. Evidently, incorporating information about previous observations leads to more efficient scheduling that optimizes coverage over the course of multiple observation rounds. 

\section{Filter Balancing}
\label{sec:filterbalancing}

If observations in multiple filters are required, \texttt{gwemopt} has the ability to implement a block-completion algorithm during the scheduling process. This means that it schedules observations in only the first filter (i.e. the first block), and then if there is time left, schedules a second pass in the next filter, and so on. This strategy minimizes the number of filter changes, which is especially advantageous since changing filters compromises observation time; the Zwicky Transient Facility (ZTF), for example, takes $\sim$\,100s to change filters with slew time taken into account \citep{Bellm2018}. 

The implementation of the block-completion algorithm may, however, lead to some challenges in scheduling observations in all requested filters for a given field. Since observations are scheduled in the second filter only after the first filter block has been completed, there will likely be a disproportionately larger number of observations in just the first filter.
This issue is pertinent to the case of ToO follow-up to GW events, as strategies for the discovery of KN counterparts \citep{AnAn2019} require observations in all requested filters to be satisfied (hence the term ``filter balancing''). This is because the characteristic rapid fading and reddening of KNe, as was seen with GW170817, can be used to identify candidates by acquiring images in at least two different filters \citep{ArHo2017,SmCh2017,PiDa2017}. The \textit{g-i} pair in particular has been shown to be most suitable in achieving this task since more KNe are expected to be detected in the \textit{i} filter relative to the others, and the combination also displays the largest color change (only second to the \textit{g-z} pair) over the days following the detection \citep{AnAn2019}.

It is important to promptly process images during the transient-filtering stage so we can narrow down the hundreds of thousands of sources of variability to a select few candidates; high-performance image subtraction pipelines have been developed for this purpose \citep[e.g.,][]{KeMa2015,GoAn2019}. In order to rule out moving objects such as near-Earth asteroids, the candidate must have a minimum of two detections separated by at least 30 minutes \citep{bellm2019scheduling}. It is justified, then, to place emphasis on scheduling at least two epochs during block scheduling\footnote{The \texttt{--doBalanceExposure} and \texttt{--doRASlices} command-line options in \texttt{gwemopt} seek to ensure this.}.

\begin{figure}[htb]
  \includegraphics[width=\columnwidth]{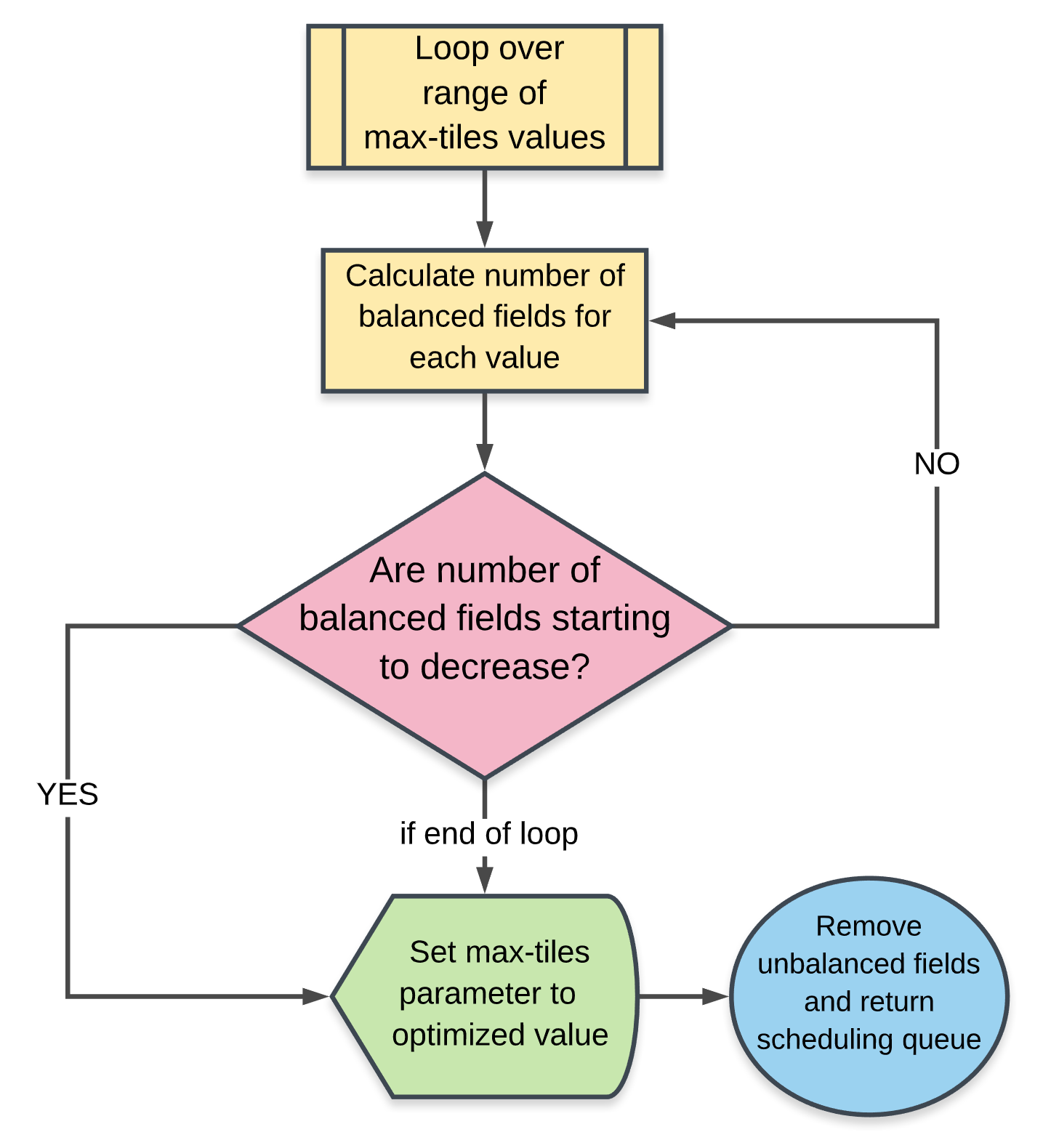}
  \caption{Step-by-step representation of the max-tiles optimization process. A ``balanced field'' is defined as a field that has all requested epochs scheduled.}
  \label{maxtilesopt}
\end{figure}
\begin{figure*}
  \begin{center}
  \includegraphics[width=0.45\textwidth]{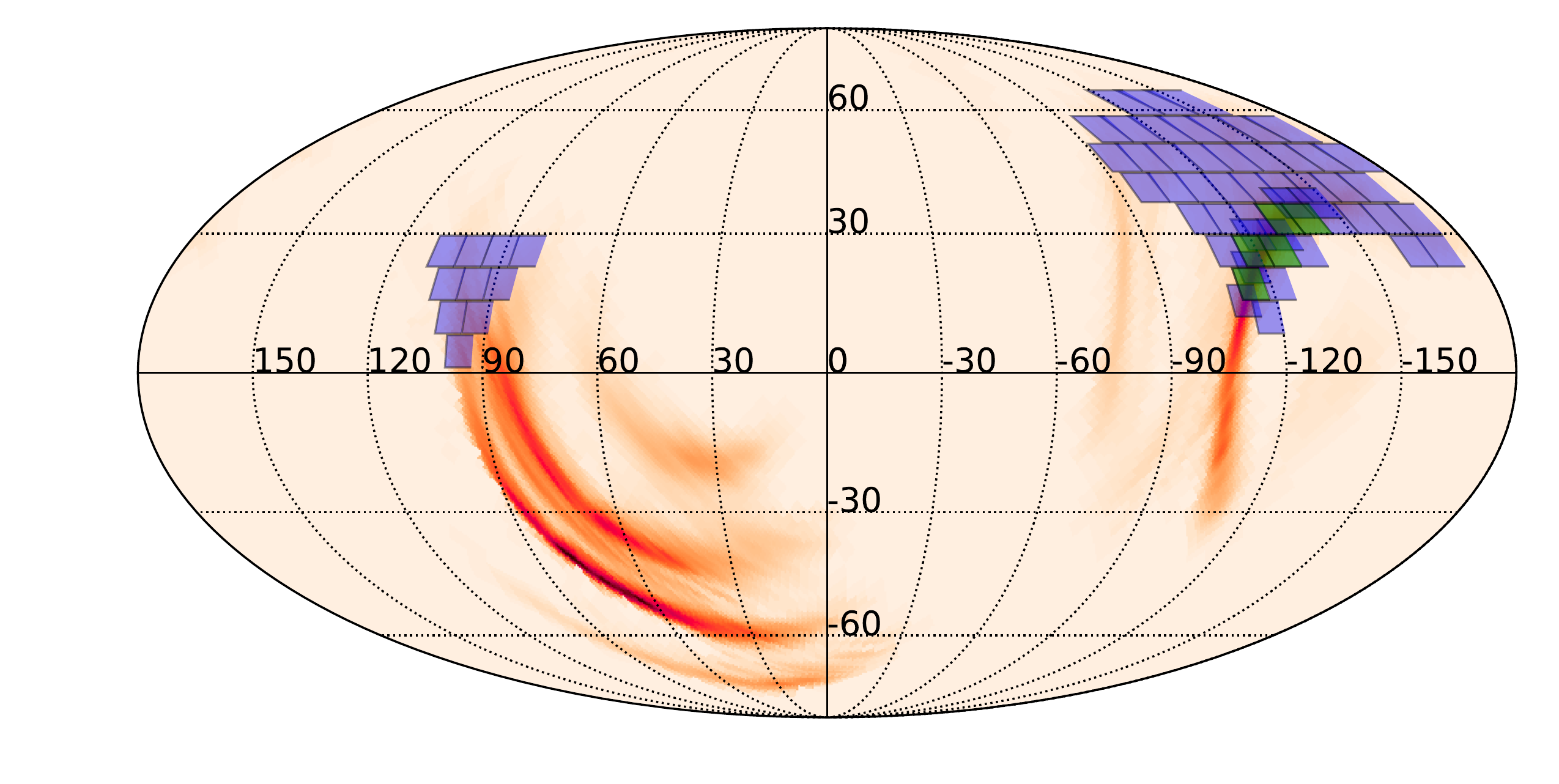}
  \includegraphics[width=0.45\textwidth]{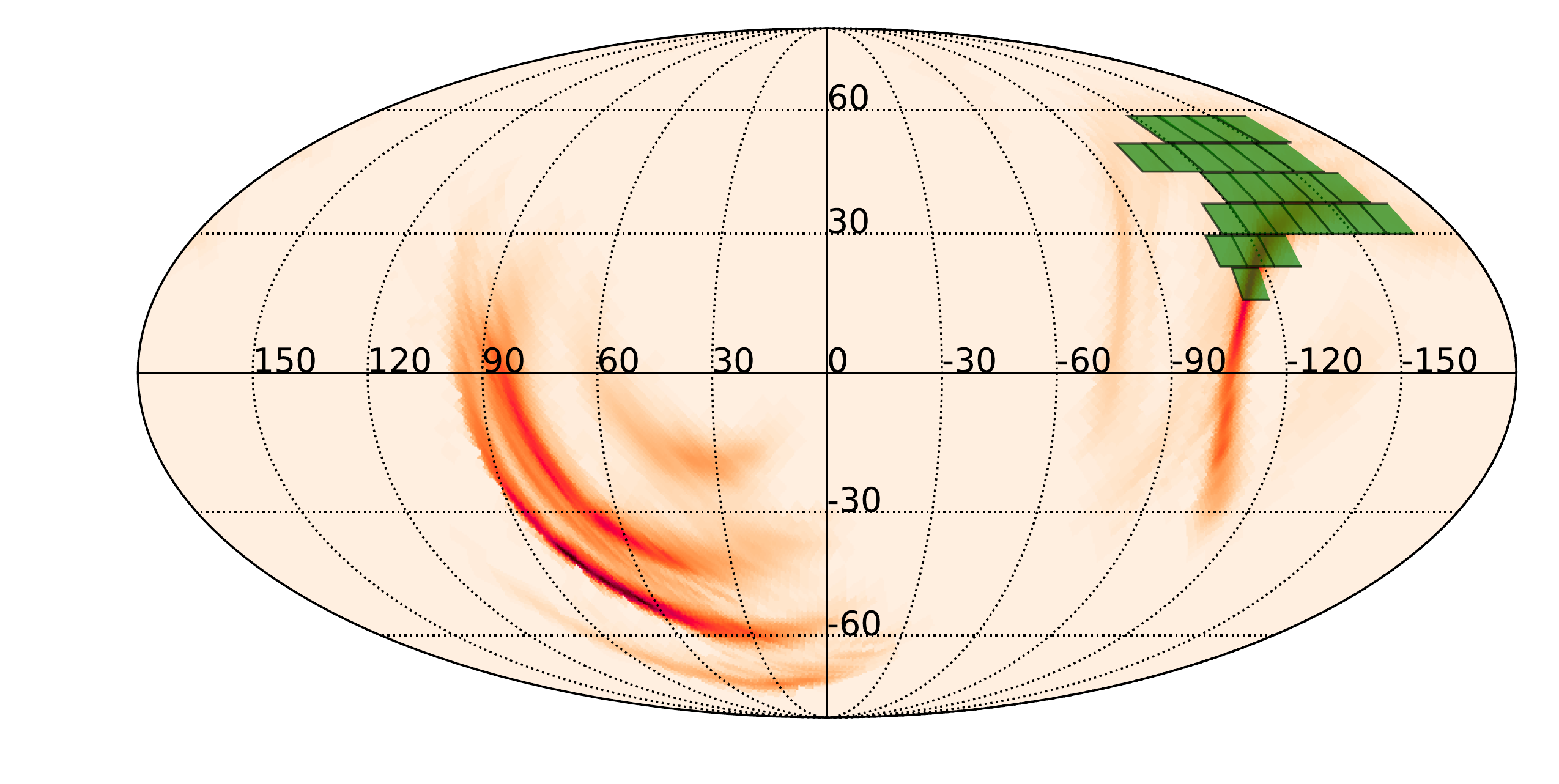}
  \includegraphics[width=0.45\textwidth]{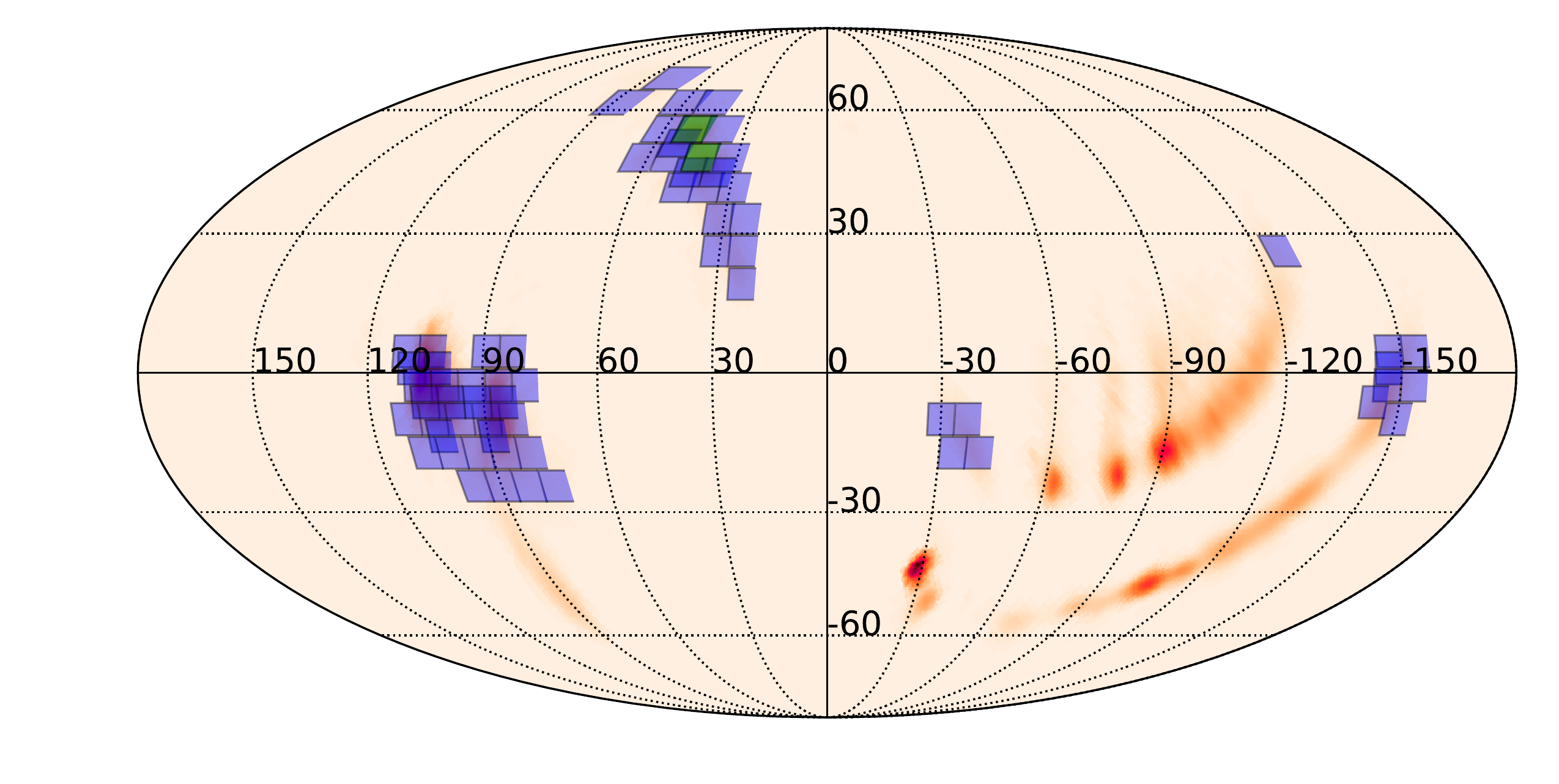}
  \includegraphics[width=0.45\textwidth]{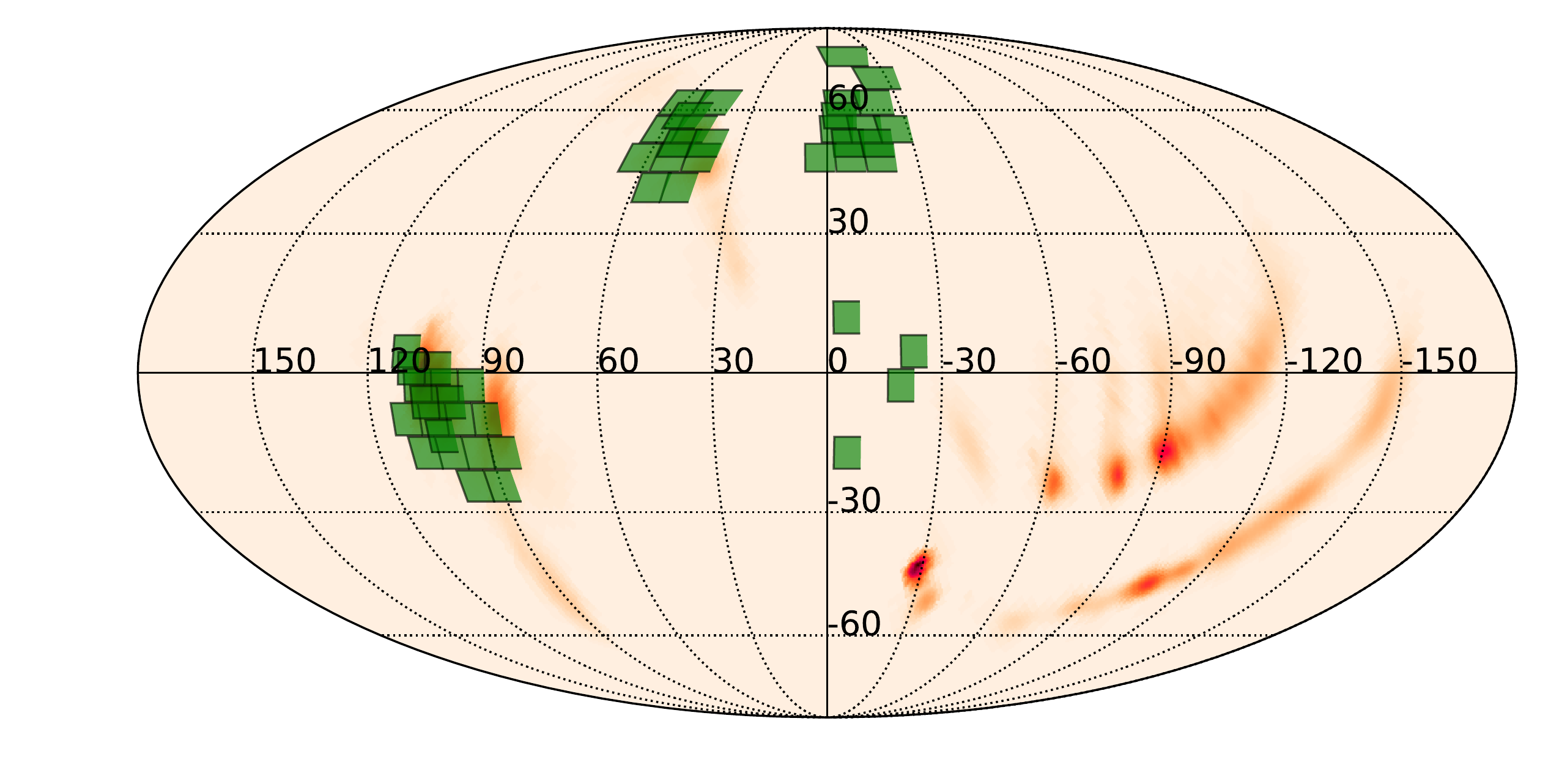}
  \end{center}
  \caption{Skymap coverage with ZTF before and after the use of the appropriate filter balancing features discussed in Section~\ref{sec:filterbalancing}. The top row displays the results for GW190425 \citep{Abbott_2020}, without (on the left) and with (on the right) the use of max-tiles optimization (Section~\ref{subsec:maxtiles}). The bottom row displays coverage for S191213g \citep{gcn26417}; in this case, we compare the results when not using any of the filter balancing features (on the left), versus when both the max-tiles optimization and right ascension slicing (Section~\ref{subsec:raslicing}) are used (on the right). Fields represented in green have had all requested observations scheduled, while those in violet have not. It is evident that the number of balanced fields increases significantly when the new filter balancing features are put to use.}
  \label{dobalancefig}
\end{figure*}

\subsection{Max-tiles optimization}
\label{subsec:maxtiles}
 Our max-tiles optimization algorithm works around the filter balancing problem by optimizing the ``max-tiles'' parameter\footnote{The corresponding command line option is \texttt{--doBalanceExposure}}, which sets an upper limit on the number of fields that are scheduled (e.g., a max-tiles value of 15 means that a maximum of 15 fields can be scheduled). It optimizes this parameter such that the number of fields with observations in all requested filters (i.e. ``balanced" fields) is maximized, iterating through a reasonable range of max-tiles values and calculating the number of balanced fields each time. If the optimization parameter starts decreasing at any point (indicating that we have reached the point where there are too many fields to ensure all required exposures are scheduled), it exits from the loop and the max-tiles parameter is now set for the rest of the scheduling process. Any scheduled fields that do not have all of the requested observations are removed before finalizing the scheduling queue. Generally, there is no limit on how many filters (repeated or otherwise) can be specified when optimizing the max-tiles value during scheduling; however, due to difficulty in revisiting the same tile multiple times without compromising coverage, it is usually optimal to limit it to around two to three epochs maximum, although this number can vary depending on the size of the localization. This process can be visualized using the flowchart in Figure~\ref{maxtilesopt}.

\subsection{Slicing in right ascension}
\label{subsec:raslicing}
Although optimizing the maximum number of tiles can help to increase the amount of balanced fields, this method only proves to be effective with certain skymaps. More specifically, in cases where the skymap contains multiple disjointed ``lobes" in the probability distribution, it is still a challenge to schedule a reasonable number of balanced fields; this is because the separation in right ascension between the different lobes leads to each lobe having its own rising and setting time. The block scheduling algorithm does not discriminate between continuous and disjointed localizations, and due to this limitation, has difficulty in scheduling both epochs within the appropriate observability windows.

We have hence implemented a feature to ``slice" the skymap in right ascension\footnote{The corresponding command line option is \texttt{--doRASlices}}, giving the scheduler the ability to distinguish between the different lobes and schedule them separately rather than treating the skymap as a whole. After slicing, the scheduler optimizes for the best order that each slice should be scheduled based on the location of the telescope. The block scheduling algorithm is still used for each slice, thus minimizing the number of filter changes; however, there are additional filter changes incorporated for the transition between each slice, which is necessary to keep up with the lobes' rising and setting times.

The results of these two features are shown in Figure~\ref{dobalancefig} for ZTF, with the left and right columns displaying the before and after skymaps. The top row displays the results for a skymap that is primarily concentrated in one area in the northern hemisphere (most of the southern lobe is not accessible), meaning that simply using the max-tiles option is sufficient. The bottom row, in turn, shows results for a skymap in which it would be useful to use both the right ascension slicing and the max-tiles option. The number of green fields (fields with all requested exposures) increases drastically in both cases, demonstrating that these two new options are effective in solving the filter balancing problem when used appropriately. More quantitatively, the cumulative probability covered (only taking into consideration tiles that have had all requested epochs scheduled) increases from 5.7\% to 11.5\% for the event shown in the top row, and from 2.1\% to 24.9\% for that shown in the bottom row.

\section{Conclusion}
\label{sec:conclusion}

In this work, we have optimized the search for GW counterparts through improvements of scheduling pipelines that rely on multi-telescope networks. We have presented the different features that we have implemented in the pursuit of making the scheduling of ToO observations more flexible and efficient, including taking previous/ongoing observations into account, and scheduling filter blocks with optimized slicing of the skymap. All of these improvements are important in addressing previous challenges associated with synoptic searches of counterparts in large and multi-lobed localizations, and work to make future electromagnetic follow-up an overall smoother and more optimally automated process.

The dynamic scheduling and filter balancing features were implemented in \texttt{gwemopt}, which is the software used to perform scheduling for both the Global Relay of Observatories
Watching Transients Happen (GROWTH) and the Global Rapid Advanced Network Devoted to the Multi-messenger Addicts (GRANDMA) projects. These networks span across multiple continents, comprising tens of observatories working in a joint effort to successfully obtain multi-wavelength observations of GW candidates. The dedicated follow-up of BNS and NSBH merger candidates undertaken by the GROWTH \citep[e.g.,][]{CoAh2019b,GoAn2019,AnGo2019,Andreoni2020} and GRANDMA \citep[e.g.,][]{GRANDMApaper,antier2020grandma} networks throughout O3 led to the realization early on that more scheduling features would need to be implemented in order to facilitate this process, prompting the subsequent development and implementation of the features described in Sections~\ref{sec:dynamicsched} and~\ref{sec:filterbalancing} throughout the rest of the observing run. The ToO marshal\footnote{\url{https://github.com/growth-astro/growth-too-marshal}} \citep{Kasliwal2018,CoAh2019} and the ICARE (\textit{Interface and Communication for Addicts of the Rapid follow-up in multi-messenger Era}) pipeline are the main drivers in coordinating the entire follow-up process for the GROWTH and GRANDMA networks respectively, and are able to do so by combining the tiling, scheduling and vetting processes into one cohesive platform. Optimizing all of the elements that lead up to the eventual classification of candidate counterparts is vital to an ultimately productive attempt at follow-up, and key to enable further progress during this exciting new era of GW astronomy.

\acknowledgments

M.~Almualla and K.~Alqassimi thank LIGO Laboratory at the California Institute of Technology for hosting the visit that led to this publication.
M.~W.~Coughlin is supported by the David and Ellen Lee Postdoctoral Fellowship at the California Institute of Technology. S.~Anand acknowledges support from the GROWTH project funded by the National Science Foundation under Grant No 1545949. Nidhal Guessoum acknowledges a research grant from the Mohammed Bin Rashid Space Centre (UAE), which supported this work.
 
\bibliographystyle{aasjournal}
\bibliography{references}

\end{document}